\documentclass[letterpaper, 10 pt, conference]{ieeeconf}  

\IEEEoverridecommandlockouts                              

\usepackage{cite}
\makeatletter
\let\NAT@parse\undefined
\makeatother
\usepackage{hyperref} 
\usepackage{xcolor}
\usepackage{graphicx}
\usepackage{amsmath}
\usepackage{amssymb}
\usepackage{tabularx}
\usepackage{multirow}
\usepackage{tikz}
\usepackage{eso-pic}
\usepackage{calc}

\title{\LARGE \bf
Economic Analysis of Smart Roadside Infrastructure Sensors for Connected and Automated Mobility
}

\author{Laurent Kloeker$^{1}$, Gregor Joeken$^{1}$ and Lutz Eckstein$^{1}$
\thanks{This research is accomplished within the project ”AUTOtech.\textit{agil}” (FKZ 01IS22088A). We acknowledge the financial support for the project by the Federal Ministry of Education and Research of Germany (BMBF).}
\thanks{$^{1}$The authors are with the research area Vehicle Intelligence \& Automated Driving, Institute for Automotive Engineering, RWTH Aachen University, 52074 Aachen, Germany
        {\tt\small laurent.kloeker@ika.rwth-aachen.de, gregor.joeken@rwth-aachen.de, lutz.eckstein@ika.rwth-aachen.de}}%
}

\begin{document}

\newcommand\copyrighttext{%
    \footnotesize \textcopyright 2023 IEEE. Personal use of this material is permitted. Permission from IEEE must be obtained for all other uses, in any current or future media, including reprinting/republishing this material for advertising or promotional purposes, creating new collective works, for resale or redistribution to servers or lists, or reuse of any copyrighted component of this work in other works. 
}

\AddToShipoutPictureBG*{%
  \AtTextLowerLeft{\raisebox{-3\baselineskip}{\makebox[\textwidth][c]{%
    \parbox{\textwidth}{\copyrighttext}}}}
}

\maketitle
\thispagestyle{empty}
\pagestyle{empty}

\begin{abstract}

Smart roadside infrastructure sensors in the form of intelligent transportation system stations (ITS-Ss) are increasingly deployed worldwide at relevant traffic nodes. The resulting digital twins of the real environment are suitable for developing and validating connected and automated driving functions and for increasing the operational safety of intelligent vehicles by providing ITS-S real-time data. However, ITS-Ss are very costly to establish and operate. The choice of sensor technology also has an impact on the overall costs as well as on the data quality. So far, there is only insufficient knowledge about the concrete expenses that arise with the construction of different ITS-S setups. Within this work, multiple modular infrastructure sensor setups are investigated with the help of a life cycle cost analysis (LCCA). Their economic efficiency, different user requirements and sensor data qualities are considered. Based on the static cost model, a Monte Carlo simulation is performed, to generate a range of possible project costs and to quantify the financial risks of implementing ITS-S projects of different scales. Due to its modularity, the calculation model is suitable for diverse applications and outputs a distinctive evaluation of the underlying cost-benefit ratio of investigated setups.

\end{abstract}

\section{INTRODUCTION}

The launch of automated and connected mobility is paving the way for a new era in transportation systems. A cornerstone of this transformation is the global augmentation of smart roadside infrastructure sensors, also known as intelligent transportation system stations (ITS-Ss), which are being installed strategically at significant traffic intersections and digital test fields.
The traffic data generated from sensor measurements can be used in various ways and are of interest to different user groups. Resulting digital twins of the real environment, for example, are suitable for the development and validation of automated driving functions \cite{Kloeker2021}. Real-time data can, in turn, be utilized by intelligent vehicles to obtain external reference information about the current traffic situation and to enhance the safety of automated driving systems. Moreover, smart city applications, like real-time traffic flow optimization, are conceivable.

In many current ITS-S applications, lidar and RGB-camera sensors are frequently used for recording the surrounding traffic \cite{Cress2021}. Despite their accuracy and versatility, establishing and operating lidar sensors is cost-intensive. The cost factor plays a significant role before setting up such installations. The use of alternative sensors, such as thermal cameras or radar sensors, will impact both costs and data quality. Consequently, the choice of sensor concept also strongly influences the addressable user groups.

Numerous studies exist for analyzing the societal benefits of intelligent infrastructure \cite{Evans2014}, \cite{Asselin2016}, \cite{Mitsakis2019}. 
Existing life cycle cost analyses (LCCAs) and cost-benefit analyses (CBAs) have so far exclusively dealt with infrastructure applications using vehicle-to-everything (V2X) communication road side units (RSUs) \cite{Rehme2019}, \cite{Degrande2021}. However, there is no such analysis for sensor-based ITS-S setups at the current time.

In this paper, we therefore present a novel modular approach to determine the costs of large-scale coverage of German traffic with ITS-Ss. We identify the initial and ongoing costs associated with the establishment and operation of a large network of ITS-Ss. We explore sensor combinations and focus on all costs associated with ITS-Ss over their lifetime to yield meaningful results.

Initially, we set several reference scenarios for the LCCA in the German context. These include the overall size of the ITS-S networks, the project lifespan, and distribution across urban, rural, and highway traffic areas. However, the model presented in this paper is not limited to these reference scenarios and can be extended to other countries with sufficient data. We introduce the \mbox{ITS-S} concepts to be analyzed and describe the mathematical modeling of the static cost model, initially without considering potential cost uncertainties. For this purpose, an analysis of the current interest rate environment and price change effects in the German market is conducted. This is followed by the calculation of capital expenditures (CapEx) and operational expenditures (OpEx). To account for financial risks, the static model is transformed into a cost model incorporating uncertainty. A Monte Carlo simulation is conducted to determine the range of project costs within the German price environment. In the final step, the expected system performance is estimated by introducing the metrics of reach, availability, and quality. The reach and availability are calculated for the reference scenario, while the quality is estimated based on a previous publication \cite{Kloeker2023} using the analytic hierarchy process (AHP). These calculated metrics are combined to form a comprehensive effectiveness indicator.

\section{RELATED WORK}\label{II}

\begin{figure*}[t]
  \includegraphics[width=\textwidth]{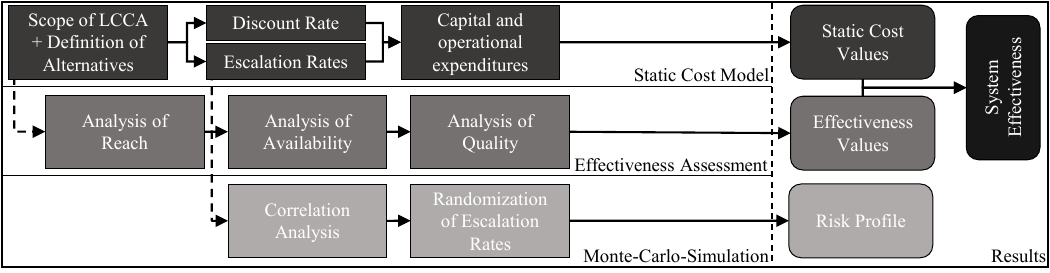}
  \caption{Method for economic analysis of smart roadside infrastructure sensors.}
  \label{fig:method}
\end{figure*}

Gao et al. \cite{Gao2017} explored differentiators between LCCAs for intelligent transportation systems (ITS) versus traditional infrastructure, citing differing inflation, uncertainty, lifecycle duration, technical obsolescence, and inventory management complexity. Noting the variation in costs and technical capabilities of ITS technologies, they proposed a key performance indicator (KPI) – system effectiveness – to make them comparable, with suggestions to quantify price change risk over time using Monte Carlo simulation.

Rehme et al. \cite{Rehme2019} investigated the cost of an RSU rollout in Dresden, Germany, via an LCCA for various technology scenarios. They considered a continuous RSU network installation, potential ITS-G5 and Cellular-V2X market penetration scenarios, and two budget restrictions. Overall, the costs were identified for five scenarios. The net present value (NPV) was used as the key assessment value. The evaluation focuses on a \textit{brownfield} premise where a lot of the required infrastructure is already in place (such as communications and power infrastructure and mounting hardware) and thus considers only the costs associated with the RSUs. A sensitivity analysis was carried out to assess cost assumption uncertainty. 

Degrande et al. \cite{Degrande2021} performed an LCCA for a highway RSU rollout in Belgium, incorporating a cost-optimal RSU placement algorithm. They used a logarithmic function to consider economies of scale and learning effects, while also performing a sensitivity analysis. Their model, based on a brownfield premise, only considered RSU's CapEx and OpEx, with an assumption of no obsolescence. 

As of our knowledge, there is no comprehensive LCCA for ITS-Ss using roadside perception sensors. This paper aims to address this gap by proposing a comprehensive LCCA framework for ITS-Ss. This framework considers price uncertainty, risk quantification via the Monte Carlo method. It also considers \textit{greenfield} situations in which no existing infrastructure is available. The suggested cost model is inherently complex due to its increased differentiation between cost positions.

\section{METHOD}\label{III}

Fig. \ref{fig:method} illustrates the foundational method employed for the economic analysis of intelligent roadside infrastructure sensors. Networks of ITS-Ss utilizing various sensor combinations, including lidar (L), radar (R), RGB cameras (C), and thermal cameras (T), are examined throughout this work. Initially, the static LCC is computed, taking into account price fluctuations in the German market, for nine German cities with diverse population sizes. Table~\ref{tab:cities} offers an overview of these cities. OpenStreetMap was utilized to determine the number of nodes necessary to encompass all intersections within the analyzed cities and their corresponding road types. Subsequently, the proposed model is converted into a dynamic cost model, and a Monte Carlo simulation is performed to quantify the financial risks associated with an ITS-S project. Following this, the method of estimating the effectiveness values for the ITS-S sensor alternatives using the AHP is described in \cite{Saaty2004}.

\begin{table}[t]
\caption{Information on Cities Selected for LCCA Based on 2021 Data}
\label{tab:cities}
\begin{center}
\vspace{-10pt}
\begin{tabularx}{86 mm}{|c||c|c|c|c|c|}
\hline
\multirow{2}{*}{\textbf{City}} & \multirow{2}{*}{\textbf{Population}} & \multicolumn{4}{c|}{\textbf{Nodes}} \\
& & \textbf{Total} & \textbf{Urban} & \textbf{Rural} & \textbf{Highway}\\
\hline
Berlin & 3,677,472 & 18,816 & 18,341 & 306 & 169\\
\hline
Hamburg & 1,853,935 & 13,358 & 12,358 & 829 & 171\\
\hline
Munich & 1,487,708 & 9,329 & 8,954 & 167 & 208\\
\hline
Cologne & 1,073,096 & 8,610 & 8,054 & 218 & 338\\
\hline
Frankfurt & 759,224 & 4,964 & 4,459 & 242 & 263\\
\hline
Duisburg & 495,152 & 5,004 & 4,680 & 172 & 152\\
\hline
Aachen & 249,070 & 2,384 & 2,077 & 230 & 77\\
\hline
Hildesheim & 100,319 & 1,456 & 1,379 & 33 & 44\\
\hline
\end{tabularx}
\end{center}
\vspace{-7pt}
\end{table}

\subsection{Static Cost Model}\label{III-A}

It is assumed that the network of ITS-S in the analyzed cities is constructed over three years and that the construction happens in the shape of a sigmoid function analogous to the work of Rehme et al \cite{Rehme2019}. The LCC are captured over a total period of 15 years and all expenditures are discounted to the NPV. 

As described in Section~\ref{II}. the LCC of an ITS project differs from conventional infrastructure projects in their inflation behavior and thus the cost assumptions need to be adjusted over time \cite{Gao2017}. To account for this the average inflation-adjusted price, changes of all major cost positions of the LCCA are captured in the form of an escalation rate \cite{Mack2012}. The escalation rates were calculated from the goods price indices published by the German Federal Statistics Office between the years 1991 and 2021 \cite{Destatis2022}. For cost items in which data was not available in this time frame, the average price change was calculated from the earliest possible data point. Since the escalation rates are inflation-adjusted the discount rate used is the real discount rate as proposed by \cite{Mack2012}. In-line with the current recommendation of the US government for carrying out LCCAs, a real discount rate of 1.75\,\% is used \cite{Usgov2022}.

The LCCA considers the CapEx and OpEx for the build-up and operation of the ITS-S network. The CapEx capture the cost of initially building up the individual ITS-Ss. All cost values are based on quotations received from hardware and service providers from a German large-scale digital test field and were broken down per ITS-S. The costs for installing each ITS-S depends on the street type (i.e. residential/urban, rural or highway) \cite{Kloeker2021}. 

In urban areas the costs of the perception sensors, V2X-RSU, local data processing computer and antennas are added up for each ITS-S. The components considered are analogous to the ones described in the ACCorD project \cite{Kloeker2021}. Even though mounting points are available in urban environments, they need to be retrofitted to allow for the installation of the ITS-Ss. The required supporting hardware such as power outlets, telematic units and various controllers are also accounted for.

On rural roads and highways, a greenfield situation is assumed, in contrast to existing literature \cite{Rehme2019, Degrande2021}. Hence, it is assumed no supporting hardware or mounting points are available on rural roads and highways. This necessitates their inclusion in the LCCA. For the construction of ITS-Ss in these areas, new network infrastructure, energy infrastructure, and mounting hardware are required \cite{Kloeker2021}. All associated hardware and installation costs are captured in the LCCA. In the ACCorD project, independent solar power systems were utilized in rural areas and highways, which is also assumed in this LCCA \cite{Kloeker2021}. To adhere to safety regulations for newly installed sensor masts, the addition of new guard rails is necessary on rural roads, and the associated costs are considered. At the end of the project duration, the value of the CapEx is prorated based on the remaining service life (RSL), which is then subtracted from the NPV. The RSL of the captured CapEx is based on the recommendations provided by the Association of German Engineers (VDI) for conducting economic analyses \cite{Vdi2012}.

In the OpEx the costs for energy, personnel, operation of a centralized back-end, renewal and maintenance are accounted for. Each node is assumed to consume up to 400\,W peak power for the local real-time data processing unit, antennas, and sensors \cite{Kloeker2021}. The energy consumption for the perception sensors varies for each setup individually. It has to be noted that this setup involves a high-performance computer with a dedicated graphics processing unit (GPU). An integrated solution would result in significantly lower power consumption but is currently unavailable for this particular application. To run the ITS-Ss and its backend, personnel is required. For the operation of the ITS-Ss network it is estimated that one employee per 250 nodes is needed. The number of employees grows proportionately with the number of ITS-Ss in smaller cities to a maximum number of 20 employees. The employees considered are skilled employees with an average market rate. For regular maintenance and inspection it is assumed that a third-party contractor is used with the costs being estimated through the rates supplied by \cite{Vdi2012}. Maintenance costs are not subject to price changes aside from inflation and are thus not adjusted further. Costs for the operation of a centralized back end to store and analyze trajectories extracted by the ITS-Ss in real-time were estimated through quotes provided by cloud computing providers. The computing power analogous to the ACCorD test field was assumed \cite{Kloeker2021} of which the cost was distributed across the ITS-Ss to get the per station cost. The renewal cost is determined by identifying the number and type of components due for replacement in each period of the LCC model. By applying the escalation rate of the individual hardware costs the new costs applicable in each time frame of the LCC is calculated.  

All CapEx and OpEx costs are added up and discounted over all considered time periods of the LCCA to determine the NPV for each sensor setup across all cities analyzed. While these are the most likely costs an ITS-S project will have based on extrapolated historical data, uncertainty in future price changes needs to be considered as described in \cite{Gao2017}.

\subsection{Monte-Carlo-Simulation}\label{III-B}

To account for uncertainty in future price changes in the model a Monte-Carlo-Simulation was carried out. To calculate the likelihood of different project costs, the escalation rates assigned to the cost positions in the model were replaced with random values. For defining the uncertainty it is not sufficient to  replace each escalation rate with a normally distributed random variable. This is because the different escalation rates such as for computers, perceptions sensors, telecommunications technology and electricity are correlating. If each variable was simulated separately without considering their correlation, the resulting spread in project costs would be smaller than is to be anticipated.

Initially, the correlation matrix of the escalation rates was computed. This analysis was conducted utilizing the monthly price fluctuations derived from data spanning February 2015 to April 2022 \cite{Destatis2022}. The starting point of February 2015 was selected as it marked the first instance wherein data was accessible for all individual components within the model. Consequently, this provided 87 data points for each of the nine escalation rates featured in the model. All correlations detected within the data set exhibited a p-value less than 0.05.

To generate random escalation rates with a similar correlation behavior as the real-world data set equation~\ref{eq:TransformationEscalationRates} is used as described in \cite{Chou2011}.
\begin{equation}
    \label{eq:TransformationEscalationRates}
    \mathbf{X} = \boldsymbol{\mu} + \mathbf{C} * \mathbf{Z}
\end{equation}

$\mathbf{X}$ is the vector of random escalation rates with the desired correlation profile over the LCCA time frame. $\boldsymbol{\mu}$ is the vector of the average escalation rates of the static model. It is transformed by adding the product of $\mathbf{C}$, which is the lower triangular matrix of the cholesky-decompostion of the covariance matrix, and $\mathbf{Z}$, which is a vector of normally distributed random variables with $\mathbf{N(0,1)}$. For each iteration of the simulation 180 vectors of randomized escalation rates are calculated, one for each month considered in the calculation. 

An additional random variable is introduced to represent the anticipated price of solid-state lidars (SSLs). Considering the likelihood of SSLs being introduced in the coming years, a reduced-price assumption has been implemented for lidar sensors within both the static and dynamic cost models. It is posited that SSLs could be integrated into the network at the midpoint of the project duration, which represents a conservative estimate. To encompass the uncertainty surrounding the future price of SSLs, a triangular distribution was assumed for the simulation's purposes. In order to generate the comprehensive ITS-S project risk profile, 100,000 iterations were executed using the described model.

\subsection{System Effectiveness}\label{III-C}

\setlength\tabcolsep{4.24pt} 
\begin{table*}[t]
\caption{LCC of Analyzed ITS-S Implementations in German Cities (BN. €)}
\label{tab:cities-LCCA}
\begin{center}
\vspace{-18pt}
\begin{tabular}{|c||c|c|c|c|c|c|c|c|c|c|c|c|c|c|c|c|c|c|c|c|}
\hline
\multirow{2}{*}{\textbf{City}} & \multicolumn{4}{c|}{\textbf{C}} & \multicolumn{4}{c|}{\textbf{R}} & \multicolumn{4}{c|}{\textbf{T}} & \multicolumn{4}{c|}{\textbf{L}} & \multicolumn{4}{c|}{\textbf{CR}}\\
& \textbf{Tot.} & \textbf{Ur.} & \textbf{Ru.} & \textbf{Hi.} & \textbf{Tot.} & \textbf{Ur.} & \textbf{Ru.} & \textbf{Hi.} & \textbf{Tot.} & \textbf{Ur.} & \textbf{Ru.} & \textbf{Hi.} & \textbf{Tot.} & \textbf{Ur.} & \textbf{Ru.} & \textbf{Hi.} & \textbf{Tot.} & \textbf{Ur.} & \textbf{Ru.} & \textbf{Hi.}\\
\hline
Berlin & 1.14 & 1.10 & 0.04 & 0.02 & 1.31 & 1.27 & 0.04 & 0.02 & 2.15 & 2.08 & 0.05 & 0.03 & 2.73 & 2.64 & 0.06 & 0.04 & 1.45 & 1.40 & 0.04 & 0.02\\
\hline
Hamburg & 0.86 & 0.76 & 0.10 & 0.02 & 0.98 & 0.88 & 0.11 & 0.02 & 1.57 & 1.42 & 0.14 & 0.03 & 1.98 & 1.81 & 0.17 & 0.04 & 1.07 & 0.97 & 0.11 & 0.02\\
\hline
Munich & 0.61 & 0.57 & 0.02 & 0.03 & 0.70 & 0.66 & 0.02 & 0.03 & 1.11 & 1.05 & 0.03 & 0.04 & 1.40 & 1.33 & 0.03 & 0.04 & 0.76 & 0.72 & 0.02 & 0.03\\
\hline
Cologne & 0.58 & 0.52 & 0.03 & 0.04 & 0.66 & 0.60 & 0.03 & 0.05 & 1.04 & 0.95 & 0.04 & 0.06 & 1.30 & 1.20 & 0.05 & 0.07 & 0.72 & 0.66 & 0.03 & 0.05\\
\hline
Frankfurt & 0.37 & 0.32 & 0.03 & 0.03 & 0.42 & 0.36 & 0.03 & 0.04 & 0.64 & 0.56 & 0.04 & 0.05 & 0.79 & 0.70 & 0.05 & 0.06 & 0.45 & 0.39 & 0.03 & 0.04\\
\hline
Duisburg & 0.37 & 0.33 & 0.02 & 0.02 & 0.41 & 0.38 & 0.02 & 0.02 & 0.63 & 0.58 & 0.03 & 0.03 & 0.79 & 0.73 & 0.04 & 0.03 & 0.45 & 0.41 & 0.02 & 0.02\\
\hline
Aachen & 0.21 & 0.18 & 0.03 & 0.01 & 0.23 & 0.20 & 0.03 & 0.01 & 0.33 & 0.29 & 0.04 & 0.01 & 0.41 & 0.35 & 0.05 & 0.02 & 0.25 & 0.21 & 0.03 & 0.01\\
\hline
Hildesheim & 0.13 & 0.12 & 0.00 & 0.01 & 0.14 & 0.13 & 0.00 & 0.01 & 0.21 & 0.19 & 0.01 & 0.01 & 0.25 & 0.23 & 0.01 & 0.01 & 0.15 & 0.14 & 0.00 & 0.01\\
\hline
\hline
\multirow{2}{*}{\textbf{City}} & \multicolumn{4}{c|}{\textbf{TR}} & \multicolumn{4}{c|}{\textbf{CL}} & \multicolumn{4}{c|}{\textbf{TL}} & \multicolumn{4}{c|}{\textbf{CRL}} & \multicolumn{4}{c|}{\textbf{TRL}}\\
& \textbf{Tot.} & \textbf{Ur.} & \textbf{Ru.} & \textbf{Hi.} & \textbf{Tot.} & \textbf{Ur.} & \textbf{Ru.} & \textbf{Hi.} & \textbf{Tot.} & \textbf{Ur.} & \textbf{Ru.} & \textbf{Hi.} & \textbf{Tot.} & \textbf{Ur.} & \textbf{Ru.} & \textbf{Hi.} & \textbf{Tot.} & \textbf{Ur.} & \textbf{Ru.} & \textbf{Hi.}\\
\hline
Berlin & 2.45 & 2.38 & 0.06 & 0.03 & 2.86 & 2.77 & 0.07 & 0.04 & 3.85 & 3.74 & 0.08 & 0.05 & 3.17 & 3.07 & 0.07 & 0.04 & 3.64 & 3.53 & 0.08 & 0.05\\
\hline
Hamburg & 1.79 & 1.63 & 0.16 & 0.03 & 2.08 & 1.89 & 0.18 & 0.04 & 2.78 & 2.54 & 0.22 & 0.05 & 2.29 & 2.10 & 0.19 & 0.04 & 2.64 & 2.41 & 0.22 & 0.05\\
\hline
Munich & 1.26 & 1.20 & 0.03 & 0.04 & 1.46 & 1.39 & 0.04 & 0.05 & 1.95 & 1.86 & 0.04 & 0.06 & 1.62 & 1.54 & 0.04 & 0.05 & 1.86 & 1.77 & 0.04 & 0.06\\
\hline
Cologne & 1.18 & 1.09 & 0.04 & 0.07 & 1.36 & 1.26 & 0.05 & 0.07 & 1.82 & 1.68 & 0.06 & 0.09 & 1.50 & 1.39 & 0.05 & 0.08 & 1.73 & 1.60 & 0.06 & 0.09\\
\hline
Frankfurt & 0.72 & 0.63 & 0.05 & 0.05 & 0.83 & 0.73 & 0.05 & 0.06 & 1.09 & 0.96 & 0.06 & 0.07 & 0.91 & 0.80 & 0.06 & 0.06 & 1.04 & 0.92 & 0.06 & 0.07\\
\hline
Duisburg & 0.72 & 0.66 & 0.03 & 0.03 & 0.82 & 0.76 & 0.04 & 0.03 & 1.09 & 1.01 & 0.05 & 0.04 & 0.91 & 0.84 & 0.04 & 0.04 & 1.04 & 0.96 & 0.05 & 0.04\\
\hline
Aachen & 0.37 & 0.32 & 0.04 & 0.02 & 0.42 & 0.37 & 0.05 & 0.02 & 0.55 & 0.48 & 0.06 & 0.02 & 0.46 & 0.40 & 0.05 & 0.02 & 0.53 & 0.46 & 0.06 & 0.02\\
\hline
Hildesheim & 0.23 & 0.21 & 0.01 & 0.01 & 0.26 & 0.24 & 0.01 & 0.01 & 0.34 & 0.32 & 0.01 & 0.01 & 0.28 & 0.27 & 0.01 & 0.01 & 0.33 & 0.31 & 0.01 & 0.01\\
\hline
\end{tabular}
\end{center}
\vspace{-7pt}
\end{table*}
\setlength\tabcolsep{6pt} 

A prominent challenge in conducting a comparative LCCA for ITS-Ss lies in the varying technical capabilities of the sensor alternatives. To address this, the system effectiveness $\mathit{SE}$ of an ITS-S setup $i \in \{1,...,n\}$ is evaluated according to equation~\ref{eq:SystemEffectiveness}, which represents the normalized ratio of the effectiveness $E$ and the LCC for each ITS-S implementation \cite{Gao2017}.
\begin{equation}
    \label{eq:SystemEffectiveness}
    \mathit{SE_i} = \frac{(\frac{E}{\mathit{LCC}})_i}{max\{(\frac{E}{\mathit{LCC}})_1, ..., (\frac{E}{\mathit{LCC}})_n\}}
\end{equation}

In the scope of this work $E$ is defined as the product of the system reach, availability, and quality and is a unitless figure. The quality is the product of the system accuracy, reliability and latency as proposed in \cite{Kloeker2023}. The reach of the ITS-Ss is the actual number of nodes in the network. For the purposes of this work the reach is not further discussed because the number of nodes is the same for all technical options analyzed. Consideration of reach is required when the implementer of an ITS-Ss project has a budget restriction which causes him to favor a more economical solution, which in turn allows the placement of more ITS-Ss in the network.

The availability metric quantifies the average duration within a 24-hour period for which the system provides data of the desired quality. In the effectiveness calculation, availability is expressed as a value ranging from 0 to 1. Primary factors influencing the availability of each sensor include lighting and weather conditions. For simplicity, the availability of a radar sensor is assumed as 1, as it remains mostly unaffected by environmental conditions. It is assumed that precipitation adversely impacts data quality for lidar and thermal cameras, resulting in the deduction of the average annual rainfall duration from their availability. To ascertain the duration of rainfall exceeding a threshold of 0.1\,mm/h, equivalent to light rain, data from 13 stations across Germany was analyzed, provided by the German Weather Service \cite{Dwd2023}. This analysis revealed that rainfall surpasses the threshold 6.88\,\% of the time in Germany, yielding availability values of 93.12\,\% for both lidar and thermal imaging cameras. For RGB cameras, we assume a reduction of this value by half, since the data quality is not or only insufficiently given in darkness.

For the quality evaluation, the AHP by Saaty is employed \cite{Saaty2004}. This choice is driven by the limited availability of real-world data concerning sensor combinations in terms of accuracy, latency, and reliability. Nevertheless, it is feasible to make verbal estimations on how the setups comparatively perform against one another. For instance, it can be confidently assumed that a lidar sensor offers higher object detection accuracy than a radar sensor. The AHP method is utilized as it facilitates the quantification of relative comparisons between setups, enabling the estimation of quality for the analyzed alternatives without necessitating costly and lengthy real-world testing. However, for an actual ITS-S project, real quality should be validated through field experiments.

To execute the AHP, four matrices were defined for pairwise comparisons. The first matrix establishes the relative preference between accuracy, latency, and reliability. For the ITS-S implementation, higher accuracy is favored as it enables additional smart city applications beyond cooperative perception, thereby providing increased utility. Moreover, it is assumed that reliability and latency are equally important, leading to the project implementer's indifference between the two.

For each sensor setup, pairwise comparisons with all alternatives concerning quality attributes are conducted. We make use of the definitions for calculating accuracy, latency, and reliability from \cite{Kloeker2023}. For accuracy, it is assumed that L$>$C$>$T$>$R, where "$>$" indicates that the L-setup is preferred over the C-setup with respect to overall sensor and object detection accuracy. For latency, T$>$C$>$R$>$L is assumed based on the conventional raw data throughput and the associated processing time and for reliability, R$>$T$>$L$>$C is assumed. The addition of extra sensors is presumed to significantly enhance overall accuracy and marginally increase system reliability. In terms of latency, extra sensors are expected to increase system latency. All matrices are tested for consistency using the consistency index, which was below 0.05 for all matrices \cite{Saaty2004}. By employing the AHP method, verbal preferences are quantified as weight values, which are then scaled in the final effectiveness calculation for easier interpretation of the results.

\section{RESULTS}
\begin{figure}[t]
  \includegraphics{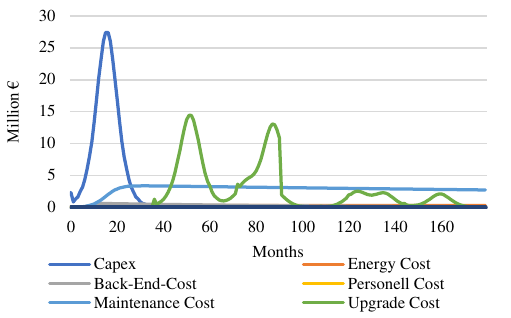}
  \caption{Cost structure of CL-setup in Cologne over a period of 15 years.}
  \vspace{-7pt}
  \label{fig:cost-structure-cologne}
\end{figure}
\begin{figure}[t]
  \includegraphics{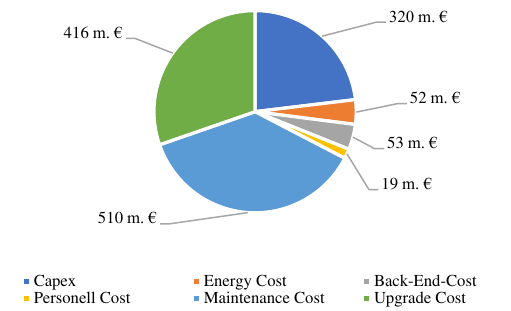}
  \caption{Cost split of CL-setup in Cologne over a period of 15 years.}
  \vspace{-9pt}
  \label{fig:cost-split-cologne}
\end{figure}
In examining the results, we primarily focus on the city of Cologne. With a population of approximately one million inhabitants, it serves as a suitable comparative indicator for other major European cities. The static analysis has revealed that the overall project cost is highly sensitive to the choice of sensors. Depending on the sensors employed, costs can range from 580 million € for a RGB camera-only implementation to 1.73 billion € for a configuration utilizing thermal camera, and radar and lidar. The LCCs for the other eight German cities, with varying population sizes and in conjunction with nine distinct sensor setups, have been computed, with the results presented in Table~\ref{tab:cities-LCCA}. It can be observed that the total costs, as well as the individual costs for all three domains - city, rural, and highway - are approximately linearly related to the city's population size. Rounding errors cannot be ruled out due to the chosen unit. Consequently, the table is well-suited for the classification and cost estimation of other European cities with similar urban structures. However, North American cities, due to their generally lower complexity in road networks, cannot be directly incorporated into this table. The same applies to Asian cities, where, assuming equal road network complexity, a higher population density prevails. Additionally, it is essential to note that equipping all intersections of a major city within the urban, rural, and highway domains with ITS-Ss in the future is not representative. This merely serves as a determination of the upper limit, which, in turn, can be linearly downscaled individually for each domain in terms of expansion levels.

Fig. \ref{fig:cost-structure-cologne} provides a detailed overview of the cost structure for the expansion of a CL setup in Cologne over a 15-year period. The focus of this analysis lies on a CL setup, as it currently reflects the most common form of ITS-S implementations. In this case, a complete expansion of nodes across all three domains is considered with 8,610 nodes in total. A wave-like pattern is observed, which results from the combination of CapEx and upgrade costs. While CapEx significantly influences costs within the first two years, the regular upgrade costs in a 4-year cycle constitute the majority of the remaining 13 years.

A cumulative summary of these costs can be found in Fig.~\ref{fig:cost-split-cologne}. CapEx account for 23\,\% of the costs, amounting to 320 million €. The remaining 77\,\%, equivalent to 1,050 million €, is attributed to operating costs. Upon examining the operating costs individually, maintenance and upgrade costs together constitute 88.2\,\% of the total. Personnel, backend, and energy costs comprise the remaining 11.8\,\%.

Table~\ref{tab:monte-carlo-simulation} presents the results of the Monte Carlo simulation for Cologne. It focuses on the C, L, and CL setups, although the analysis was also conducted for all other configurations. The findings indicate that the standard deviation increases in tandem with elevated sensor costs. This is particularly evident for setups incorporating lidar sensors, given the added uncertainty surrounding the future price of SSLs. The standard deviation for the CL setup amounts to 36.39 million €, with a range of 299.95 million € between the upper and lower 5\,\% quartiles. The relatively high standard deviation and spread, in comparison to other sensor setups, imply that the additional financial risks relative to alternative configurations must be taken into account. Consequently, municipalities or businesses implementing ITS-S projects need to allocate provisions for substantial cost increases under unfavorable market conditions.

The system effectiveness values for the different setups in the cologne region are shown in Table~\ref{tab:system-effectiveness-cologne}. The best results for quality, availability $\mathit{E}$ and $\mathit{LLC}$ are highlighted in bold. Based on the result of the AHP the CR setup brings the best effectiveness to cost ratio. The second best option identified was the CL setup. Even though the CR setup provides significantly less accuracy than the CL for example, it also costs 48.79\,\% less which still gives it a good overall system effectiveness if there is a significant budget restriction in place. The third best option determined by using the AHP is the CRL setup. This is because the implementation of the extra Radar provides a boost in reliability of the system and enables the supply of data even in adverse conditions.
\begin{table}[t]
\caption{Monte Carlo Results for LCC in Cologne (M. €)}
\label{tab:monte-carlo-simulation}
\begin{center}
\vspace{-10pt}
\begin{tabular}{|c||c|c|c|}
\hline
\textbf{Value} & \textbf{C} & \textbf{L} & \textbf{CL}\\
\hline
Average & 573.12 & 1,308.54 & 1,369.97\\
\hline
Standard deviation & 9.70 & 34.68 & 36.39\\
\hline
Median & 572.84 & 1,307.63 & 1,368.90\\
\hline
Upper 5\% quartile & 589.85 & 1,370.81 & 1,435.18\\
\hline
Lower 5\% quartile & 557.63 & 1,253.26 & 1,312.02\\
\hline
Range & 89.99 & 294.64 & 299.95\\
\hline
\end{tabular}
\end{center}
\vspace{-7pt}
\end{table}
\setlength\tabcolsep{1pt} 
\begin{table}[t]
\caption{System Effectiveness Values for ITS-Ss Implementations in Cologne}
\label{tab:system-effectiveness-cologne}
\begin{center}
\vspace{-18pt}
\begin{tabular}{|c||c|c|c|c|c|c|c|c|c|}
\hline
\textbf{Cologne} & \textbf{C} & \textbf{R} & \textbf{T} & \textbf{L} & \textbf{CR} & \textbf{TR} & \textbf{CL} & \textbf{TL} & \textbf{CRL}\\
\hline
Quality [-] & 0.079 & 0.050 & 0.074 & 0.110 & 0.098 & 0.079 & 0.181 & 0.136 & \textbf{0.193}\\
\hline
Availability [-] & 0.466 & \textbf{1.000} & 0.931 & 0.931 & 0.740 & 0.975 & 0.748 & 0.969 & 0.748\\
\hline
Reach [-] & 8610 & 8610 & 8610 & 8610 & 8610 & 8610 & 8610 & 8610 & 8610\\
\hline
$\mathit{E}$ [-] & 316.5 & 431.1 & 595.7 & 880.4 & 621.5 & 666.7 & 1165.1 & 1132.0 & \textbf{1244.9}\\
\hline
$\mathit{LCC}$ [bn. €] & \textbf{0.58} & 0.66 & 1.04 & 1.30 & 0.72 & 1.18 & 1.36 & 1.82 & 1.50\\
\hline
\hline
$\mathbf{SE}$ \textbf{[-]} & 0.63 & 0.76 & 0.66 & 0.78 & \textbf{1.00} & 0.65 & 0.99 & 0.72 & 0.96\\
\hline
\end{tabular}
\end{center}
\vspace{-7pt}
\end{table}
\setlength\tabcolsep{6pt} 

\section{CONCLUSIONS}

In this paper, a comprehensive LCCA was conducted to determine the total costs of an extensive implementation of ITS-Ss for various German cities. The developed mathematical model is characterized by a high level of detail and takes into account the temporal structure of costs. Furthermore, the analysis incorporated changes in the prices of individual components over time. Based on historical data, the real price changes for all components were estimated.

Through both the LCCA and a Monte-Carlo simulation, it was demonstrated that the selection of sensor technology has a significant influence on the resulting total costs and project risk. Depending on the specific implementation of ITS-S applications, different costs can be expected. The installation of ITS-Ss with cameras and lidars at all intersections of a German example city with about one million inhabitants (e.g. Cologne) causes a LCC in the range of 1.312 billion € to 1.435 billion €. While it is not representative that all intersections in the urban, rural, and highway domains of a major city will be equipped with ITS-Ss in the future, this value provides a good indication for further estimation of smaller scales.

The modularity of the calculation approach enables the determination of LCC for various scenarios and scales. Thus, the calculation model contributes to well-founded decision-making and cost estimation in the implementation and financing of ITS-S applications and is therefore suitable for different users with various downstream functions. These users include, in particular, cities, municipalities, and other private and public institutions that plan to implement future ITS-S applications in the context of automated and connected mobility, smart cities, and digital twins.






\bibliographystyle{IEEEtran}
\bibliography{references}

\end{document}